\documentstyle[12pt,epsfig]{article}
\title{ Cosmic ray drift, the second knee and galactic anisotropies}
\author{Juli\'an Candia$^a$, Silvia Mollerach$^b$ and Esteban Roulet$^b$\\
$^a${\it Departamento de F\'{\i}sica, Universidad Nacional de La Plata, 
CC67,}\\{\it La Plata 1900, Argentina}\\
$^b$ {\it CONICET, Centro At\'omico Bariloche, Av. Bustillo 9500,}\\
{\it Bariloche 8400, Argentina}}

\begin{document}
\maketitle

\begin{abstract}
We show that the second knee in the cosmic ray spectrum (i.e. the
steepening occurring at $E\simeq 4\times 10^{17}$~eV) could be related
to drift effects affecting the heaviest nuclear component, the iron
group nuclei, in a scenario in which the knee at $3\times  10^{15}$~eV
indicates the onset of drift effects in the lighter proton component. 
We also study the anisotropies resulting from diffusion and drift
currents in the Galaxy, showing their potential relevance to account
for the AGASA observations at $E\sim 10^{18}$~eV, before the
extragalactic component becomes dominant.
\end{abstract}

There are a couple of observations regarding cosmic rays (CRs) at energies
$\sim10^{18}$~eV, i.e. just below the ankle, which still remain
puzzling (see e.g. \cite{na00,yo98}). 
One is the steepening in the spectrum taking place at an
energy $E_{sk}\simeq 4\times10^{17}$~eV, which is most apparent
in the Yakutsk \cite{af95} and Fly's Eye stereo \cite{bi93,bi94} data, 
and is usually referred to as the `second knee', at 
which the spectral index $\alpha$ (such that $dN/dE \propto
E^{-\alpha}$) changes from $\alpha \simeq 3$ below $E_{sk}$ to 
$\alpha \simeq 3.3$ above $E_{sk}$, although probably in a smooth
way. The other is the $4\%$ anisotropy at $E \simeq 0.8-2 \times 
10^{18}$~eV reported by the AGASA collaboration \cite{ha99}, 
with an excess observed from a
direction near the galactic center and a (probably associated) deficit
near the galactic anticenter direction. Also some excess of events in
the direction of the Cygnus region (along the Orion spiral arm) has
been observed. An excess from a direction near the galactic center was 
also found in the SUGAR data \cite{be01}, but it has a smaller spatial
extension (consistent with a point-like source) and its direction is
few degrees away from the AGASA reported excess.
The fact that these anisotropies are associated to galactic features is
quite important, because it is believed that the galactic cosmic ray
component should just be fading away at $E>10^{18}$eV, since the CR
spectrum should become dominated by the extragalactic component above the
ankle, which lies at energies $\sim 5\times 10^{18}$eV. Hence, the
energies at which these features take place are probably the highest
ones at which the CR spectrum is still dominated by the galactic
component, and the features observed may actually be related with the
very same process responsible for the fading away of this component.

In this work we want to show that the 
steepening of the spectrum at $E_{sk}$ can be directly related 
 to the enhancement of the escape
mechanism of the iron group nuclei due to efficient drift effects, and
that the observed anisotropies can be linked to the overall CR motion arising
from these drifts. It is
interesting that these processes would then be directly related to the
scenario  in which 
 the knee in the spectrum at $E_{knee}\simeq 3\times 10^{15}$~eV
results just from the transition from a regime in which the CR transport 
is dominated by transverse diffusion to one dominated by the Hall
diffusion, i.e. by drift effects \cite{pt93,ca02}. 
This scenario is based on the fact that each CR component of charge $Z$
starts to be affected by drifts at an energy $E\simeq Z~E_{knee}$, and its
spectrum progressively steepens, with the spectral index finally
changing by $\Delta \alpha \simeq 2/3$ in a decade of energy. The
envelope of the total spectrum obtained by adding together the
different nuclear components nicely fits the change from a spectrum
$\propto E^{-2.7}$ below the knee, to one $\propto E^{-3}$ above
it. However, above $10^{17}$~eV, where all the lighter components are
strongly suppressed, the dominant iron component will progressively
steepen its spectrum  until the overall spectrum becomes $\propto
E^{-3.3}$ above a few $\times 10^{17}$~eV, hence
reproducing the behavior observed at the second knee. This scenario
also naturally explains the transition towards a
heavier composition above the knee reported by several experiments.

The change in the spectrum from the diffusion to the drift dominated
regimes can be simply understood from the steady state diffusion
equation ${\bf \nabla}\cdot{\bf J} = Q$, where $Q$ is the source and
the CR current is related to the CR density $N$ through
\begin{equation}
{\bf J}=-D_\perp {\bf \nabla}_\perp N- D_{\parallel} 
{\bf \nabla}_{\parallel} N
+D_A {\bf b}\times{\bf \nabla} N,
\label{current}
\end{equation}
with ${\bf b}$ being the unit vector in the direction of the regular
magnetic field ${\bf B}_{reg}$, i.e. ${\bf b}\equiv 
{\bf B}_{reg}/|{\bf B}_{reg}|$, and 
${\bf \nabla}_\parallel={\bf b}({\bf b}\cdot
{\bf \nabla})$, while ${\bf \nabla}_\perp={\bf \nabla}-
{\bf \nabla}_\parallel$. The components of the diffusion tensor are 
$D_{\parallel}$ (along the direction of ${\bf B}_{reg}$), 
$D_\perp$ in the
perpendicular direction, while $D_A$ is associated to the
antisymmetric part and determines the drift effects. Assuming for
simplicity that the regular magnetic field is directed in the
azimuthal direction and that in a first approach the Galaxy can be
considered to have cylindrical symmetry, one finds that  $D_\parallel$
plays no role in the diffusion equation. Using now the fact that
(for a Kolmogorov spectrum of fluctuations in the turbulent magnetic
field component)
$D_\perp ({\bf x})\simeq D_\perp^0 ({\bf x}) (E/E_0)^{1/3}$, while
$D_A({\bf x}) \simeq D_A^0 ({\bf x}) E/E_0$ \cite{pt93}, one finds that at low energies
($E<Z~E_{knee}$) the transverse diffusion is the dominant process
affecting the transport of CRs and
leading to $dN/dE \propto (D_\perp^0/D_\perp) dQ/dE \propto
E^{-\beta-1/3}$, where $\beta$ is the spectral index of the source,
while at high energies ($E\gg Z~E_{knee}$) one has instead 
$dN/dE \propto (D_A^0/D_A) dQ/dE \propto E^{-\beta-1}$.

\begin{figure}
\centerline{{\epsfxsize=4truein \epsffile{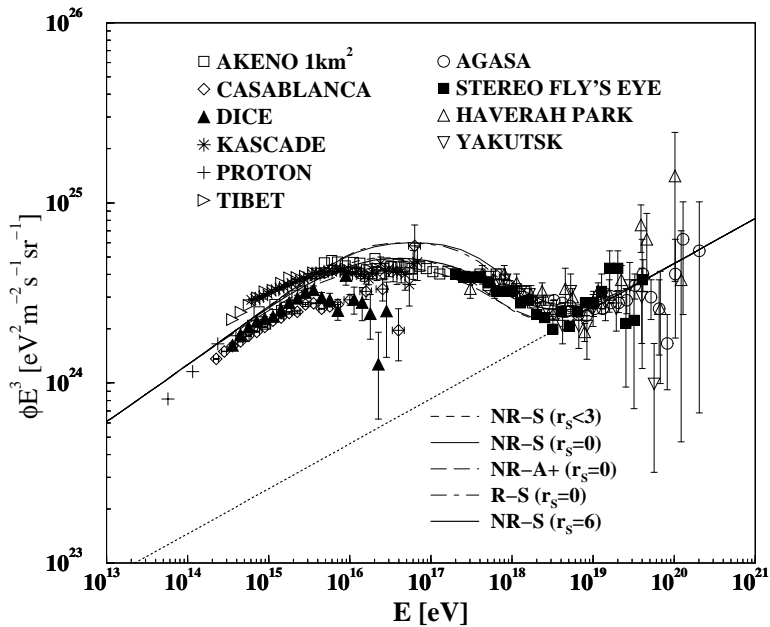}}}
\caption{CR spectra computed for different galactic magnetic field
models and source locations, which correspond to the same
source and model parameters as in Figures 3 and 4
(see more details in the text). 
The dotted straight line is the extragalactic flux from ref.~[1].
Also shown are the relevant experimental data points.}
\label{fig1}
\end{figure}

In Figure 1 we display the CR spectrum, computed following ref. 
\cite{ca02}  for
the galactic component, and adding an extragalactic component $\propto
E^{-2.75}$ (with normalization taken from \cite{na00}) 
to fit the observations beyond the ankle. Different curves
represent different assumptions for the source distribution and for
the galactic magnetic field model. The adopted galactic source spectra are
featureless  extrapolations of the spectra measured below the knee,
and hence the resulting changes in slope are just due to the energy
dependence of the diffusion process responsible for the escape of CRs
from the Galaxy. The choice of the model parameters was just guided by
the requirement of correctly reproducing the first knee, and they are
all quite plausible. As is apparent from the figure, the fit to the
data is in general remarkably good.

Also notice that if the extragalactic flux can be considered to be
isotropic, it will not be  enhanced by the Galactic diffusion process 
if no reacceleration in the Galaxy occurs. This may be understood as
being a consequence of the Liouville theorem \cite{cl96}, and can also be seen
from the diffusion equations by noting that the solution in the
absence of sources and with isotropic boundary conditions at the
outskirts of the Galaxy is just to have a constant density, so that there
is no way in which an overdensity in the extragalactic flux could 
arise from diffusion processes alone.

\begin{figure}
\centerline{{\epsfxsize=3truein \epsffile{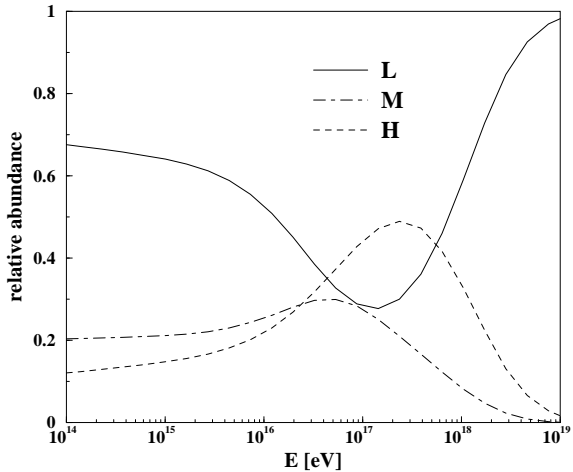}}}
\caption{Fractional abundances of different CR components, 
assuming that the extragalactic flux consists mainly of
protons. The elements are grouped into light $(1\leq Z\leq 5)$,
medium $(6\leq Z\leq 19)$ and heavy $(20\leq Z\leq 26)$ components. The data 
correspond to the same source and magnetic field model as 
in Figure~3.}
\label{fig2}
\end{figure}

The relative abundances of different elements vs. energy are plotted
in Figure~2, assuming that the extragalactic flux consists mainly of
protons. One obtains that in our model the CR composition at $E\simeq
5\times 10^{17}$~eV
consists of about 50$\%$ of galactic iron group nuclei (mostly Fe and
Mn), of a proportion ($\sim 20\%$)
of intermediate mass galactic nuclei ($Z=6$--19) while the rest
consists  essentially
of galactic and extragalactic protons. These numbers are quite consistent with the Haverah Park
determination of nuclear abundances using the lateral distribution of
air showers \cite{av02}, which suggest that in the range
$2\times 10^{17}$--$10^{18}$~eV only $\sim 30$\% of the CRs are light, with
the rest being mostly heavy nuclear species.
 Also, the composition change to a lighter mix 
starting around the energy of the second knee \cite{ab01}
is explained in this scenario
as iron group nuclei drift away from  the galactic plane with increasing
efficiency for increasing energies. 

Turning now to consider the anisotropies resulting in this scenario,
these will be produced by the galactic cosmic ray component alone, since as
just discussed, under the assumption that at $10^{18}$~eV the
extragalactic component is originally isotropic it will remain so
after the propagation effects in the Galaxy  are taken into
account. The anisotropy associated to the galactic component is given
by \cite{be90}
\begin{equation}
 \delta=\frac{3~{\bf J}}{c~N},
\end{equation} 
where $N$ is the total (galactic plus extragalactic) CR density, and 
${\bf J}$ is the CR current introduced in Eq. (\ref{current}).

Under the assumption of cylindrical symmetry and that the regular
magnetic field is in the azimuthal direction, one has that 
${\bf \nabla}_\perp N={\bf \nabla} N$ and that the CR current will be
perpendicular to the regular magnetic field (i.e. lying on the $r-z$
plane). The contribution to the CR current arising from the transverse 
diffusion will 
be in the direction of ${\bf \nabla} N$ (and hence orthogonal to the
isodensity contours), while the drift part will be orthogonal to ${\bf
\nabla} N$ (and hence parallel to the isodensity contours). Under
these simplifying assumptions, the components of the anisotropy will
then read
\begin{eqnarray}
\delta_r=\frac{3}{c~N}\left(-D_\perp\frac{\partial N}{\partial r}+
D_A {\rm sign}(B_{reg}^\phi)\frac{\partial N}{\partial z}\right) ,\\
\delta_z=\frac{3}{c~N}\left(-D_\perp\frac{\partial N}{\partial z}-
D_A {\rm sign}(B_{reg}^\phi)\frac{\partial N}{\partial r}\right).
\end{eqnarray} 

\begin{figure}[t]
\centerline{{\epsfxsize=4.5truein \epsffile{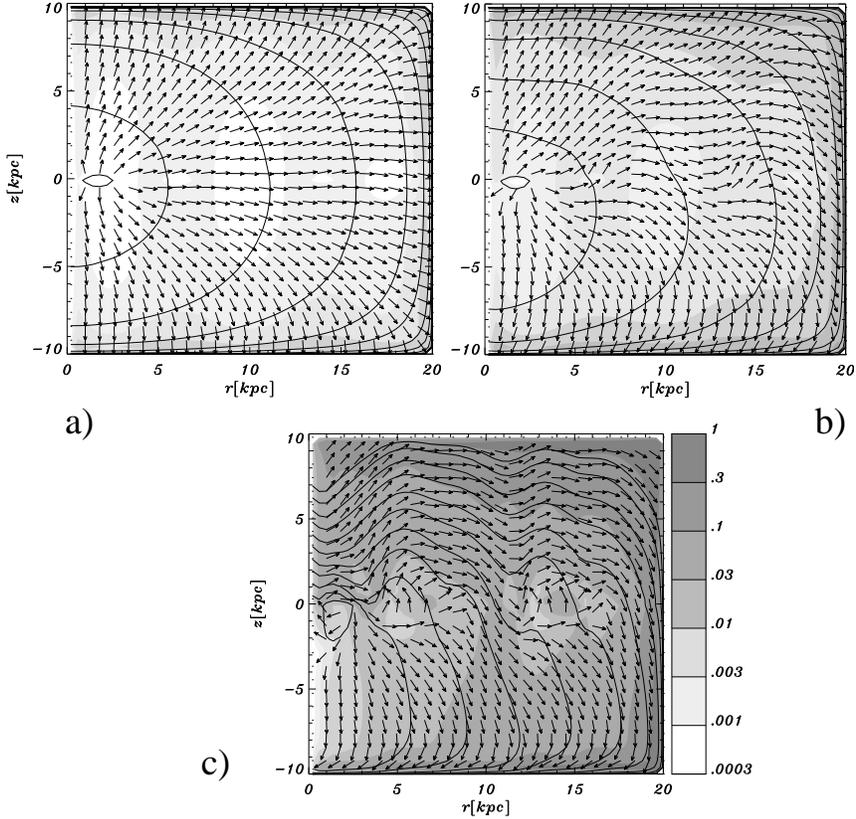}}}
\caption{Isodensity contours for different CR energies, with a)
corresponding to $E/Z=10^{14}$~eV, b) to $E/Z=E_{knee}$ and c) to
$E/Z=1.5\times 10^{18}$~eV/26. Every two contours there is a change by an order
of magnitude in the densities. The arrows represent the direction of
the anisotropy vectors, while the shadings indicate the amplitude of
the anisotropies. The model corresponds to a source in the galactic
plane constant inside 3~kpc, 
and the magnetic field structure described in the text.}
\label{fig3}
\end{figure}

In Figures 3 and 4 we display, for different galactic magnetic field and
source models, the density contours resulting from the numerical
integration of the diffusion equations (solid lines) as well as the
direction of the anisotropy vectors and with different shadings the
amplitudes of the anisotropies. The results are for given $E/Z$
values, i.e. for different energies according to which nuclear species
are considered. Our location corresponds to $r_0=8.5$~kpc and $z_0\simeq 0$.
The relative importance of the diffusion and drift
components in the different galactic locations can be directly
inferred from the relative directions between the isodensity contours and
the arrows. Figure~3.a corresponds to energies well below the first
knee ($E/Z=10^{14}$~eV), where the diffusion is completely dominated
by $D_\perp$. The source is assumed to be in the galactic plane and to
have a constant strength  for $r<3$~kpc, while vanishing at larger
radii. We
see that the resulting density contours are symmetric with respect to
the galactic plane, and the anisotropies are perpendicular to them and
quite small ($\sim 10^{-4}$). Figure~3.b is for an energy
corresponding to $E/Z\simeq E_{knee}$, where the drift effects start to
be non-negligible and the local anisotropies are $\sim 10^{-3}$, 
while figure~3.c corresponds to rigidities such
that $E/Z\simeq 1.5\times 10^{18}$~eV/26, displaying the behavior that iron
nuclei will follow for energies near $1.5\times 10^{18}$~eV. In this regime the
densities are largely determined by the drifts, and they reflect the
general asymmetries of the regular magnetic field at large scales. The
local anisotropies are in this case of order $10^{-2}$.

Figure 3 was done for a regular magnetic field consisting of a disk
component with reversals between arms, a vertical scale height
$z_d=0.5$~kpc and a local strength $B_0^{disk}=-0.75\ \mu$G$\hat\phi$. The
halo regular component had no reversals, a scale height $z_h=5$~kpc,
was symmetric with respect to the galactic plane and had a local value
of $B_0^{halo}=-0.75\ \mu$G~$\hat\phi$. The random component had a
similar scale height, $z_r=5$~kpc and a local rms value of $B_0^{rand}=1.5\
\mu$G (see ref.~\cite{ca02} for details of the magnetic field profiles
adopted). In figure~4 we display results equivalent to those of
figure~3.c but for different galactic magnetic field parameters and
geometries  
and  for different source
locations $r_s$. 
For instance, $NR$ corresponds to a halo without reversals, while $R$
(bottom left panel) corresponds to a halo with reversals similar to
those adopted for the disk model. The label $S$($A$) stands for
symmetric (antisymmetric) halo models, i.e. such that 
$B_0^{halo}(r,z) =B_0^{halo}(r,-z)$ ($B_0^{halo}(r,z)
=-B_0^{halo}(r,-z)$), while the $+$ sign in the $A+$ label (top right
panel) is because sign$(B_0^{halo}\cdot B_0^{disk})$ is positive at our
galactocentric radius for $z>0$.
Clearly the CR densities, the drifts and the
corresponding anisotropies vary a lot from model to model, even if all
models provide acceptable fits to the local CR spectrum, and hence it
is seen that the observed local anisotropies can provide a useful
handle to establish the general properties of the galactic magnetic field. 

\begin{figure}
\centerline{{\epsfxsize=5.5truein \epsffile{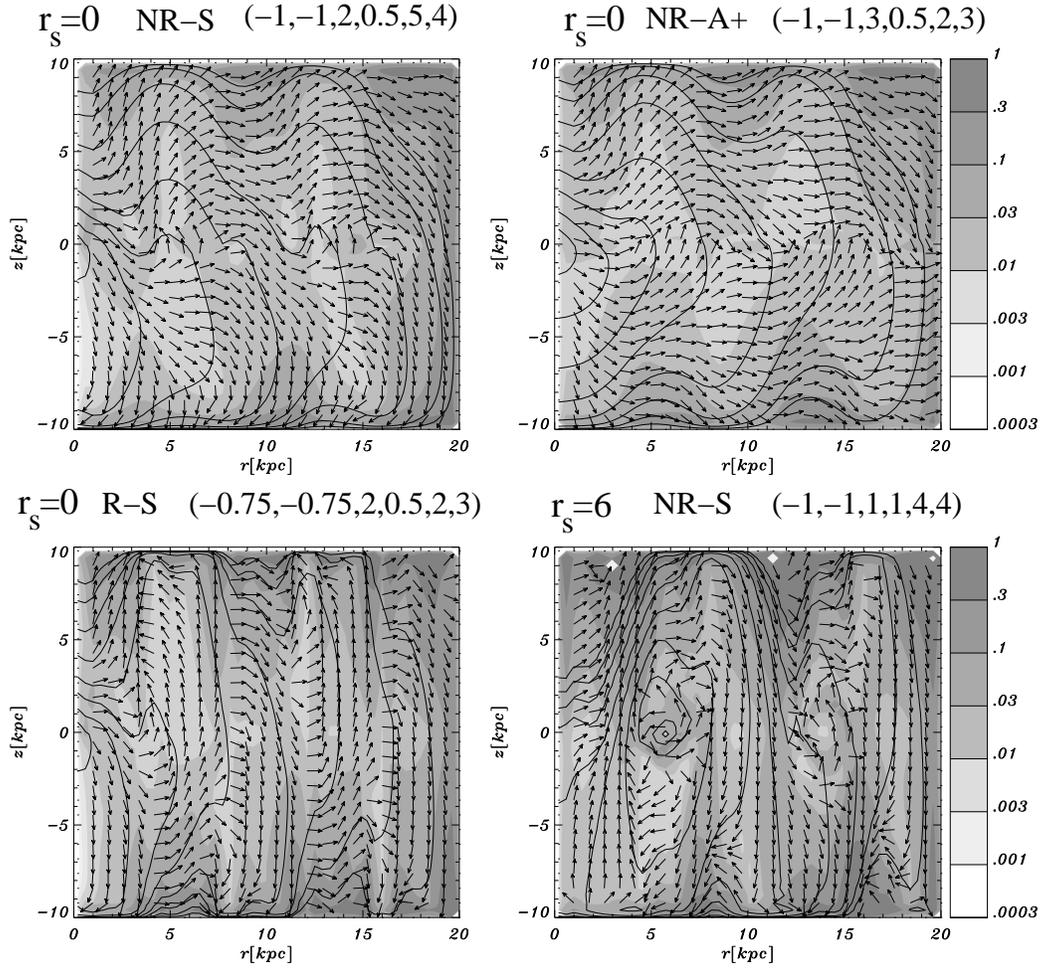}}}
\caption{CR densities and drifts for different magnetic field models
and source locations. In parenthesis are indicated the values of
$(B_0^{disk},B_0^{halo},B_0^{rand},z_d,z_h,z_r)$ in $\mu$G and kpc
respectively.} 
\label{fig4}
\end{figure}

It is important to notice that both the galactic center excess and
the galactic anticenter deficit observed by AGASA may be reproduced by an 
overall diffusion flow of the CRs near our location 
in the direction of the galactic
anticenter, with an amplitude leading to  a $4\%$ anisotropy \cite{te01}. 
Since one should have reached at the relevant energies 
 the regime in which the drift
becomes dominant with respect to the transverse diffusion, and using
the fact that locally sign$(B_{reg}^\phi)=-1$, one should have as an
approximate estimation that
\begin{equation}
\delta_r\simeq -\frac{3~D_A}{c~N}\frac{\partial N}{\partial z}\ \ ,\ \ 
\delta_z\simeq \frac{3~D_A}{c~N}\frac{\partial N}{\partial r}.
\end{equation}
The overall local asymmetry will then point approximately in the
 radial direction if locally $|\partial N/\partial z| \gg
|\partial N/\partial r|$. Its amplitude can be estimated 
from the expression of the antisymmetric diffusion coefficient, which is
\begin{equation}
D_A\simeq \frac{c r_L}{3}\frac{(\omega\tau_A)^2}{1+(\omega\tau_A)^2},
\label{da}
\end{equation}
where  $r_L$ is the CR Larmor
radius (with $\omega\simeq c/r_L$ the corresponding gyrofrequency)
and $\tau_A$ the
relevant time scale of velocity decorrelations (see ref. \cite{ca02}
and references therein).
This leads to 
\begin{equation}
|\delta_r|\simeq \frac{r_L}{h}\frac{(\omega\tau_A)^2}{1+(\omega\tau_A)^2}
\simeq 
0.04~\frac{E}{10^{18}{\rm eV}}\frac{26}{Z}\frac{\rm kpc}{h}
\frac{\mu {\rm G}}{|{\bf B}_{reg}|}
\frac{(\omega\tau_A)^2}{1+(\omega\tau_A)^2} ,
\label{aest}
\end{equation}
where $h\equiv |\partial \ln N/\partial z|^{-1}$ is the local 
vertical scale height of variation of the CR density.
Let us notice that due to the mix of different nuclear species
present, one has that actually $\delta=\sum_if_i\delta_i$, with $f_i$
being the fractional abundances of galactic species of charge $Z_i$
(plotted in figure~2), while $\delta_i$ their corresponding local
anisotropy at the energy under consideration. Although the presence of
an extragalactic proton component tends to reduce the fractions $f_i$,
the nuclei lighter than Fe tend to have larger values of $\delta_i$
(for some fixed energy), having then the tendency of increasing the value of
$\delta$. As a result,  the values displayed in figures 3.c and 4,
which are the anisotropies of Fe nuclei for $1.5\times 10^{18}$~eV, are indeed 
good estimators of the total anisotropy $\delta$ at this energy.

 We see from Eq.~(\ref{aest})  that the anisotropies  steadily grow with
energy, reaching a few $\%$ at $10^{18}$eV if the typical scale height
of the CR density variations are of order a kpc. 
Significant vertical gradients in the CR density on the galactic plane
naturally result in symmetric magnetic field models, 
since the vertical drifts in those
models are similarly oriented in both hemispheres. To have the
asymmetry vector  pointing away from the galactic center clearly
requires the local value of $\partial N/\partial z$ to be negative, something
which generally results when CRs have the tendency to drift to
negative values of $z$, and this happens more
pronouncedly e.g. in models where a symmetric halo field is pointed in
the $-\hat\phi$ direction. Also, in order that locally $|\partial N/\partial
r|<|\partial N/\partial z|$ it is convenient that the
source be not too close to our location. 
The model $NR-A+$ (top right) also shows the interesting feature that
drifts tend to converge towards the galactic plane (the opposite would
occur in $A-$ models), resulting in an outward drift for $z\simeq
0$. On the contrary, a halo model following the reversals of the disk
(or a disk model alone without a regular halo), has very pronounced
vertical drifts (bottom left panel). The factor $(\omega\tau_A)^2/
(1+(\omega\tau_A)^2)$ in Eqs.~(\ref{da}) and (\ref{aest})
 reflects the suppression
in $D_A$ in the presence of strong turbulence. It approaches unity if
$B^{rand}<B^{reg}$ but it can become small if the  turbulence
is large, and this would then suppress the anisotropies. 

Let us mention here that other attempts have been made to explain the
galactic center excess as resulting from CRs arriving almost
rectilinearly from a central source \cite{cl00,me01,be02}. Even if the
CR were protons, the deflection produced by the regular (and random)
galactic magnetic fields will largely exceed the $10^\circ$
displacement between the observed excess and the galactic center
direction (to produce this small deflection the source should be
closer than $\sim 2$~kpc \cite{me01}). Hence, the hypothesis that the
excess is due to neutrons produced from accelerated protons or nuclei
through photopion production or spallation processes around the
sources in the center of the galaxy has been explored as an
alternative explanation \cite{ha99,me01}. The interesting fact about
this proposal is that the decay length of neutrons is just $\sim
10$~kpc~$(E/10^{18}$~eV), being of the order of the galactocentric
distance for the energies relevant for the anisotropies, and hence
this would also explain why no significant 
 excess is seen at lower energies. Anyhow, to
have such a powerful neutron source is not particularly
natural\footnote{On the contrary, the maximum energy of galactic
protons in our scenario needs not be larger that $\sim 10^{17}$~eV,
a value which is clearly much more plausible.}, and
furthermore there is no way in which such scenarios can also account
for the deficit observed in the anticenter direction.

As regards the excess of events observed from the Cygnus direction,
let us notice that a similar explanation to that discussed here can in principle be searched
in terms of an overall CR diffusion along the spiral arm. This would
clearly require a departure from the assumption of cylindrical
symmetry and to take into account the parallel diffusion of CRs along
the arms. A general prediction of this kind of scenario would be that
a deficit associated to the opposite direction would result, something
of potential interest for future observations in the southern
hemisphere with the AUGER observatory. 

\section*{Acknowledgments}
Work supported by CONICET and Fundaci\'on Antorchas, 
Argentina.

\end{document}